\documentstyle[12pt]{article}
\begin{document}
\begin{center}
{\large \bf Light--Cone Current Algebra, $\pi^0$ Decay, and $e^+ e^-$
Annihilation}
\end{center}
\bigskip
\bigskip
{\bf W.A. Bardeen}\\
{\bf H. Fritzsch}\\
{\bf M. Gell--Mann}\\
\\
\\
1. Introduction\\
2. Light--cone algebra\\
3. Statistics and alternative schemes\\
4. Derivation of the $\pi^0 \rightarrow 2 \gamma $ amplitude in the
PCAC approximation\\
\\
\\
{\bf I. Introduction}\\
{\small The indication from deep inelastic electron scattering experiments
at SLAC that Bjorken scaling may really hold has motivated an extension of
the hypotheses of current algebra to what may be called light--cone current
algebra.$^1$ As before, one starts from a field theoretical quark model
(say one with neutral vector ``gluons'') and abstracts exact algebraic
results, postulating their validity for the real world of hadrons. In
light--cone algebra, we abstract the most singular term near the light cone in
the commutator of two--vector or axial vector currents, which turns out
to be given in terms of bilocal current operators that reduce to local
currents when the two space--time points coincide. The algebraic properties
of these bilocal operators, as abstracted from the model, give a number of
predictions for the Bjorken functions in deep inelastic electron and
neutrino experiments. None is in disagreement with experiment. These
algebraic properties, by the way, are the same as in the free quark model.\\
\\
From the mathematical point of view, the new abstractions differ from the
older ones of current algebra (commutators of ``good components'' of
current densities at equal times or on a light plane) in being true only
formally in a model with interactions, while failing to each order of
renormalized perturbation theory, like the scaling itself. Obviously it
is hoped that, if the scaling works in the real world, so do the relations
of light--cone current algebra, in spite of the lack of cooperation from
renormalized perturbation theory in the model.\\
\\
The applications to deep inelastic scattering involve assumptions only
about the connected part of each current commutator. We may ask whether the
disconnected part -- for example, the vacuum expected value of the
commutator of currents -- also behaves in the light--cone limit as it
does formally in the quark--gluon model, namely, the same as for a free
quark model. Does the commutator of two currents, sandwiched between the
hadron vacuum state and itself, act at high momenta exactly as it would
for free quark theory? If so, then we can predict immediately and trivially
the high--energy limit of the ratio
\begin{displaymath}
\sigma \left( e^+ e^- \rightarrow \, \, {\rm hadrons} \, \,
\right)/ \sigma \left( e^+ + e^- \rightarrow \mu^+ + \mu^- \right)
\end{displaymath}

for one--photon annihilation.\\
In contrast to the situation for the connected part and deep inelastic
scattering, the annihilation results depend on the statistics of the
quarks in the model. For three Fermi--Dirac quarks, the ratio would be
$\frac{2}{3}^2 + \left( - \frac{1}{3}^2 \right)^2 + \left( - \frac{1}{3}
\right)^2 = \frac{2}{3}$, but do we want Fermi--Dirac quarks? The
relativistic ``current quarks'' in the model, which are essentially
square roots of currents, are of course not identical with ``constituent
quarks'' of the na\''ive, approximate quark picture of baryon and meson
spectra. Nevertheless, there should be a transformation, perhaps even a
unitary transformation, linking constituent quarks and current quarks (in
a more abstract language, a transformation connecting the symmetry group
$\left[ SU(3) \times SU(3) \right]_{W, \infty, \, \, \, {\rm strong}}$
of the constituent quark picture of baryons and mesons, a subgroup of
$\left[ SU(6)\right]_{W, \infty, \, \, \, {\rm strong}} \, ^2$ with
the symmetry group $\left[ SU(3) \times SU(3) \right]_{W, \infty,
\, \, \, {\rm currents}} \, ^3$ generated by the vector and axial
vector charges). This transformation should certainly preserve quark
statistics. Therefore the indications from the constitutent quark picture
that quarks obey peculiar statistics should suggest the same beha\-vior
for the current quarks in the underlying relativistic model from which
we abstract the vacuum behavior of the light--cone current
commutator.$^4$\\
\\
In the constituent quark picture of baryons,$^5$ the ground--state wave
function is described by ({\bf 56,1}), $L = 0^+$ with respect to
$\left[ SU(6) \times 0(2) \times SU(6) \times SU(3) \right]$ or ({\bf 56},
$L_z = 0$) with respect to $\left[ SU(6) \times O(2) \right]_W$. It is
totally symmetric in spin
and $SU(3)$. In accordance with the simplicity of the picture, one
might expect the space wave function of the ground state to be totally
symmetric. The entire wave function is then symmetrical. Yet baryons are
to be antisymmetrized with respect to one another, since they do obey
the Pauli principle. Thus the peculiar statistics suggested for quarks
has then symmetrized in sets of three and otherwise antisymmetrized. This
can be described in various equivalent ways. One is to consider
``para--Fermi statistics of rank $3^{''6}$ and then to impose the
restriction that all physical particles be fermions or bosons; the quarks
are then ficticious (i.e. always bound) and all physical three--quark
systems are totally symmetric overall. An equivalent description, easier
to follow, involves introducing nine types of quarks, that is, the usual
three types in each of three ``colors,'' say red, white, and blue. The
restriction is then imposed that all physical states and all observable
quantities like the currents be singlets with respect to the $SU(3)$ of
color (i.e., the symmetry that manipulates the color index). Again, the
quarks are fictitious. Let us refer to this type of statistics as
``quark statistics.''\\
\\
If we take quark statistics seriously and apply to current quarks as well
as constituent quarks, then the closed--loop processes in the models
are multiplied by a factor of 3, and the asymptotic ratio $\sigma 
\left( e^+ e^- \rightarrow \, \, \, \, {\rm hadrons} \, \right)$
$/ \sigma \left( e^+ e^- \rightarrow \mu^+ \mu^- \right)$ becomes
$3 \cdot \frac{2}{3} = 2$.\\
\\
Experiments at present are too low in energy and not accurate enough to
test this prediction, but in the next year or two the situation should
change. Meanwhile, is there any supporting evidence? Assuming that the
connected light--cone algebra is right, we should like to know whether
we can abstract the disconnected part as well, and whether the statistics
are right. In fact, there is evidence from the decay of the $\pi^0$ into
$2 \gamma $. It is well known that in the partially conserved axial current
(PCAC) limit, with $m_{\pi}^2 \rightarrow 0$, Adler and others$^7$ have
given an exact formula for the decay amplitude $\pi^0 \rightarrow 2 \gamma$
in a ``quark--gluon'' model theory. The amplitude is a known constant
times $\left( \sum Q_{1/2} \, ^2 - \sum Q_{- 1/2} \, ^2 \right)$, where
the sum is over the types of quarks and the charges $Q_{1/2}$ are those of
$I_z = \frac{1}{2}$ quarks, while the charges $Q_{-1/2}$ are those of
$I_z = - \frac{1}{2}$ quarks. The amplitude agrees with experiment, within
the errors, in both sign and magnitude if
$\sum Q_{1/2} \, ^2 - \sum Q_{- 1/2} \, ^2 = 1$.$^8$ If we had three
Fermi--Dirac quarks, we would have $\left( \frac{2}{3} \right)^2 -
\left( - \frac{1}{3} \right)^2 = \frac{1}{3}$, and the decay rate would be
wrong by a factor of $\frac{1}{9}$. With ``quark statistics,'', we get
$\frac{1}{3} \cdot 3 = 1$ and everything is all right, assuming that
PCAC is applicable.\\
\\
There is, however, the problem of the derivation of the Adler formula. In
the original derivation a renormalized perturbation expansion is applied
to the ``quark--gluon'' model theory, and it is shown that only the
lowest--order closed--loop diagram survives in the PCAC limit,$^9$ so
that an exact expression can be given for the decay amplitude. Clearly
this derivation does not directly suit our purposes, since our light--cone
algebra is not obtainable by renormalized perturbation theory term by term.
Of course, the situation might change if all orders are summed.\\
\\
Recently it has become clear that the formula can be derived without
direct reference to renormalized perturbation theory, from considerations
of light--cone current algebra. Crewther has contributed greatly to
clarifying this point,$^{10}$ using earlier work of Wilson$^{11}$ and
Schreier.$^{12}$ Our objectives in this chapter are to call attention to
Crewther's work, to sketch a derivation that is somewhat simpler than
his, and to clarify the question of statistics.\\
\\
We assume the connected light--cone algebra, and we make the further
abstraction, from free quark theory or formal ``quark-gluon'' theory,
of the principle that not only commutators but also products and
physically ordered products of current operators obey scale invariance
near the light cone, so that, apart from possible subtraction terms
involving four--dimensional $\delta $ functions, current products near the
light cone are given by the same formula as current commutators, with the
singular functions changed from $\varepsilon \left( z_0 \right) \delta
\left[ \left( z^2 \right) \right]$ to $\left( z^2 - i \varepsilon z_0
\right)^{-1}$ for ordinary products or $\left( z^2 - i \varepsilon
\right)^{-1}$ for ordered products.\\
\\
Then it can be shown from consistency arguments that the only possible
form for the disconnected parts (two--, three--, and four--point
functions) is that given by free quark theory or formal ``quark--gluon''
theory, with only the coefficient needing to be determined by
abstraction from a model. (In general, of course, the coefficient could
be zero, thus changing the physics completely.) Then, from the light--cone
behavior of current products, including connected and disconnected parts,
the Adler formula for $\pi^0 \rightarrow 2 \gamma $ in the PCAC limit
can be derived in terms of that coefficient.\\
\\
If we take the coefficient from the model with ``quark statistics,''
predicting the asymptotic ratio of $\sigma \left( e^+ e^- \rightarrow
\, \, {\rm hadrons} \right) / \sigma
\left( e^+ e^- \rightarrow \mu^+ \mu^-
\right)$ to be 2 for one--photon annihilation, we obtain the correct
value of the $\pi^0 \rightarrow 2 \gamma$ decay amplitude, agreeing with
experiment in magnitude and sign. Conversely, if for any reason we do not
like to appeal to the model, we can take the coefficient from the
observed $\pi^0 \rightarrow 2 \gamma $ amplitude and predict in that way
that the asymptotic value of $\sigma \left( e^+ e^- \rightarrow \, \,
{\rm hadrons} \right) / \sigma \left( e^+ e^- \rightarrow \mu^+ \mu^-
\right)$ should be about 2.\\
\\
Some more complicated and less attractive models that agree with the
observed $\pi^0 \rightarrow 2 \gamma $ amplitude are discussed in Section
3.\\
\\
\\
2. {\bf LIGHT--CONE ALGEBRA}\\
\\
The ideas of current algebra stem essentially from the attempt to abstract,
from field theoretic quark models with interactions, certain algebraic
relations obeyed by weak and electromagnetic currents to all orders in the
strong interaction and to postulate these relations for the system of
real hadrons, while suggesting possible experimental tests of their
validity. In four dimensions, with spinor fields involved, the only
renormalizable models are ones that are barely renormalizable, such as
a model of spinors coupled to a neutral vector ``gluon'' field. Until
recently, the relations abstracted, such as the equal-time commutation
relations of vector and axial charges or charge densities, were true in
each order of renormalized perturbation theory in such a model. Now,
however, one is considering the abstraction of results that are true only
formally, with canonical manipulation of operators, and that fail, by
powers of logarithmic factors, in each order of renormalized
perturbation theory, in all barely renormalizable models (although
they might be all right in a super--renormalizable model, if there were
one).\\
\\
The reason for the recent trend is, of course, the tendency of the deep
inelastic electron scattering experiments at SLAC to encourage belief in
Bjorken scaling, which fails to every order of renormalized perturbation
theory in barely renormalizable models. There is also the availability
of beautiful algebraic results, with Bjorken scaling as one of their
predictions, if formal abstractions are accepted. The simplest such
abstraction is that of the formula giving the leading singularity on the
light cone of the connected part of the commutator of the vector or axial
vector currents,$^1$ for example:
\begin{eqnarray}
\left[ F_{i \mu} (x), F_{j \nu} (y) \right] & \hat{=} & \left[ F_{i \mu}
\, ^5 (x), F_{j \nu} \, ^5 (y) \right] \nonumber \\ \nonumber \\
& \hat{=} & \frac{1}{4 \pi} \partial_{\rho} \left\{ \varepsilon
\left( x_0 - y_0 \right) \delta \left[ (x - y)^2 \right] \right\}
\nonumber \\  \nonumber \\
& & \times \left\{ \left( if_{ijk} - d_{ijk} \right)
\left[ s_{\mu \nu \rho \sigma} F_{k \sigma} (y, x) + i
\varepsilon_{\mu \nu \rho \sigma} F_{k \sigma} \, ^5 (y, x)
\right) \right] \nonumber \\   \nonumber \\
& & \left. + \left( if_{ijk} + d_{ijk} \right)
\left[ s_{\mu \nu \rho \sigma}
F_{k \sigma} (x, y) - i \varepsilon_{\mu \nu \rho \sigma} F_{k \sigma} \, ^5
(x, y)  \right] \right\}
\end{eqnarray}

On the right--hand side we have the connected parts of bilocal operators
$F_{i \mu} (x, y)$ and $F_{i \mu} \, ^5 (x, y)$, which reduce to the local
currents $F_{i \mu} (x)$ and $F_{i \mu} \, ^5 (x)$ as $x \rightarrow y$.
The bilocal operators are defined as observable quantities only in the
vicinity of the light--cone, $(x - y)^2 = 0$. Here
\begin{equation}
s_{\mu \nu \rho \sigma} = \delta_{\mu \rho} \delta_{\nu \sigma} +
\delta_{\nu \rho} \delta_{\mu \sigma} - \delta_{\mu \nu}
\delta_{\rho \sigma} \, .
\end{equation}

Formula 1 gives Bjorken scaling by virtue of the finite matrix elements
assumed for $F_{i \mu} (x , y)$ and $F_{i \mu} \, ^5 (x, y)$; in fact,
the Fourier transform of the matrix element of $F_{i \mu} (x, y)$ is just
the Bjorken scaling function. The fact that all charged fields in the
model have spin $\frac{1}{2}$ determines the algebraic structure of the
formula and gives the prediction $\left( \sigma_L / \sigma_T \right)
\rightarrow 0$ for deep inelastic electron scattering, not in contradiction
with experiment. The electrical and weak charges of the quarks in the
model determine the coefficients in the formula, and give rise to numerous
sum rules and inequalities for the SLAC--MIT experiments in the Bjorken
limit, again none in contradiction with experiment.\\
\\
The formula for the leading light--cone singularity in the commutator
contains, of course, the physical information that near the light cone
we have full symmetry with respect to $SU(3) \times SU(3)$ and with
respect to scale transformations in coordinate space. Thus there is
conservation of dimension in the formula, with each current having
$l = - 3$ and the singular function $x - y$ also having $l = - 3$.\\
\\
A simple generalization of the abstraction that we have considered
turns into a closed sy\-stem, called the basic light--cone algebra. Here
we commute the bilocal operators as well, for instance,
$F_{i \mu} (x, u)$ with $F_{j \nu} (y, v)$, as all of the six intervals
among the four space--time points approach $0$, so that all four points
tend to lie on a lightlike straight line in Minkowski space. Abstraction
from the model gives us, on the right--hand side, a singular function of
one coordinate difference, say $x - v$, times a bilocal current $F_{ia}$
or $F_{ia} \, ^5$ at the other two points, say $y$ and $u$, plus an
expression with $(x, \nu)$ and $(y, u)$ interchanged, and the system
closes algebraically. The formulas are just like Eq. 1. We shall assume
here the validity of the basic light--cone algebraic system, and discuss
the possible generalization to products and to disconnected parts. In
Section 4, we conclude from the generalization to products that the form
of an expression like $< \, {\rm vac} \mid F_{ia} (x) F_{j \beta}
(y, z) \mid \, \, {\rm vac} \, >$ for disconnected parts is uniquely
determined from the consistency of the connected light--cone algebra to
be a number $N$ times the corresponding expression for three free
Fermi--Dirac quarks, when $x, y$, and $z$ tend to lie on a straight
lightlike line. The $\pi^0 \rightarrow 2 \gamma$ amplitude in the PCAC
approximation is then calculated in terms of $N$ and is proportional to
it. Thus we do not want $N$ to be zero.\\
\\
The asymptotic ratio
$\sigma \left( e^+ e^- \rightarrow \, \, {\rm hadrons} \right) / \sigma
\left( e^+ e^- \rightarrow \mu^+ \mu^- \right)$ from one--photon
annihilation is also proportional to $N$. We may either determine $N$
from the observed $\pi^0 \rightarrow 2 \gamma $ amplitude and then
compute this asymptotic ratio approximately, or else appeal to a model
and abstract the exact value of $N$, from which we calculate the
amplitude of $\pi^0 \rightarrow 2 \gamma $. In a model, $N$ depends on
the statistics of the quarks, which we discuss in the next section.\\
\\
\\
{\bf 3. STATISTICS AND ALTERNATIVE SCHEMES}\\
\\
As we remarked in Section 1, the presumably unwanted Fermi--Dirac
statistics for the quarks, with $N = 1$, would give
$\sigma (e^+ e^- \rightarrow \, \, {\rm hadrons} /\sigma
\left( e^+ e^- \rightarrow \mu^+ \mu^- \right) \rightarrow 2/3)$.
(Such quarks could be real particles, if necessary.) Now let us consider
the case of ``quark statistics,'' equivalent to para--Fermi statistics
or rank 3 with the restriction that all physical particles be bosons
or fermions. (Quarks are then fictitious, permanently bound. Even if we
applied the restriction only to baryons and mesons, quarks would still be
fictitious, as we can see by applying the principle of cluster
decomposition of the S--matrix.)\\
\\
The quark field theory model or the ``quark--gluon'' model is set up
with three fields, $q_R$, $q_B$, and $q_W$, each with three ordinary
$SU(3)$ components, making nine in all. Without loss of generality, they
may be taken to anticommute with one another as well as with themselves.
The currents all have the form $\bar q_R q_R + \bar q_B q_B +
\bar q_W q_W$, and are
singlets with respect to the $SU(3)$ of color. The physical states too are
restricted to be singlets under the color $SU(3)$. For example, the
$q \bar q$ configuration for baryons is only
$\bar q_R q_R + \bar q_B q_B + \bar q_W q_W$, and the $qqq$ configuration
for bayons is only
$q_R q_B q_W - q_B q_R q_W + q_W q_R q_B - q_R q_W q_B + q_B q_W q_R -
q_W q_B q_R$.
Likewise all the higher configurations for baryons and mesons are required
to be color singlets.\\
\\
We do not know how to incorporate such restrictions on physical states into
the formalism of the ``quark--gluon'' field theory model. We assume without
proof that the asymptotic light--cone results for current commutators
and multiple commutators are not altered. Since the currents are all
color singlets, there is no obvious contradiction.\\
\\
The use of quark statistics then gives $N = 3$ and $\sigma \left( e^+ e^-
\rightarrow \, \, {\rm hadrons} \, \, \right) / \sigma \left( e^+ e^-
\rightarrow \mu^+ \mu^- \right) \rightarrow 2$. This is the value that we
predict.\\
\\
We should, however, examine other possible schemes. First, we might
treat actual para--Fermi statistics of rank 3 for the quarks without
any further restriction on the physical states. In that case, there are
excited baryons that are not fermions and are not totally symmetric in the
$3q$ configuration; there are also excited mesons that are not bosons.
Whether the quarks can be real in this case without violating the
principle of ``cluster decomposition'' (factorizing of the $S$--matrix
when a physical system is split into very distant subsystems) is a matter
of controversy; probably they cannot. In this situation, $N$ is
presumably still $3$.\\
\\
Another situation with $N = 3$ is that of a physical color $SU(3)$ that
can really be excited by the strong interaction. Excited baryons now exist
that are in octets, decimets, and so on with respect to color, and mesons
in octets and higher configurations. Many conserved quantum numbers exist,
and new interactions may have to be introduced to violate them. This is a
wildly speculative scheme. Here the nine quarks can be real if necessary,
that is, capable of being produced singly or doubly at finite energies
and identified in the laboratory.\\
\\
We may consider a still more complicated situation in which the
relationsship of the physical currents to the current nonet in the
connected algebra is somewhat modified, namely, the Han--Nambu
scheme.$^{13}$ Here there are nine quarks, capable of being real, but
they do not have the regular quark charges. Instead, the $u$ quarks have
charges 1,1,0, averaging to $\frac{2}{3}$; the $d$ quarks have charges
0,0, -1, averaging to $- \frac{1}{3}$; and the $s$ quarks also have
charges 0,0, -1, averaging to $- \frac{1}{3}$. In this scheme, not
only can the analog of the color variable really be excited, but also
it is excited even by the electromagnetic current, which is no longer a
``color'' singlet. Since the expressions for the electromagnetic current
in terms of the current operators in the connected algebra are modified,
this situation cannot be described by a value of $N$. It is clear,
however, from the quark charges, that the asymptotic behavior of the
disconnected part gives, in the Han--Nambu scheme, $\sigma \left( e^+ e^-
\rightarrow \, \, {\rm hadrons} \, \, \right) / \sigma \left( e^+ e^- 
\rightarrow \mu^+ \mu^- \right) \rightarrow 4$. Because the formulas
for the physical currents are changed, numerical predictions for deep
inelastic scattering are altered too. For example, instead of the
inequality $\frac{1}{4} \le \left[ F^{en} (\xi) / F^{ep} (\xi) \right]
\le 4$ for
deep inelastic scattering of electrons from neutrons and protons, we would
have $\frac{1}{2} \le \left[ F^{en} (\xi) / F^{ep} (\xi) \right] \le 2$.
However,
comparison of asymptotic values with experiment in this case may not be
realistic at the energies now being explored. The electromagnetic
current is not a color singlet; it directly excites the new quantum
numbers, and presumably the asymptotic formulas do not become applicable
until above the thresholds for the new kinds of particles. Thus, unless and
untill entirely new phenomena are detected, the Han--Nambu scheme really
has little predictive power.\\
\\
A final case to be mentioned is one in which we have ordinary ``quark
statistics'' but the usual group $SU(3)$ is enlarged to $SU(4)$ to
accomodate a ``charmed'' quark $u'$ with charge $\frac{2}{3}$ which has
not isotopic spin or ordinary strangeness but does have a nonzero value
of a new conserved quantum number, charm, which would be violated by
weak interactions (in such a way as to remove the strangeness--changing
part from the commutator of the hadronic weak charge operator with its
Hermitian conjugate). Again the expression for the physical currents in
terms of our connected algebra is altered, and again the asymptotic value
of $\sigma \left( e^+ e^- \rightarrow \, \, {\rm hardons} \right) / \sigma
\left( e^+ e^- \rightarrow \mu^+ \mu ^- \right)$ is changed, this time to
$\left[ \left( \frac{2}{3} \right)^2 + \left( - \frac{1}{3} \right)^2
+ \left( - \frac{1}{3} \right)^2 + \left( \frac{2}{3} \right)^2  \right]
\cdot 3 = \frac{10}{3}$. Just as in the Han--Nambu scheme, the
predictive power is very low here until the energy is above the threshold
for making ``charmed'' particles.\\
\\
We pointed out in Section 1 that for three Fermi--Dirac quarks the Adler
amplitude is too small by a factor of 3. For all the other schemes
quoted above, however, it comes out just right and the decay amplitude
of $\pi^0 \rightarrow 2 \gamma $ in the PCAC limit agrees with
experiment. One may verify that for all of these schemes
$\sum Q_{1/2} \, ^2 - \sum Q_{-1/2} \, ^2 = 1$. The various schemes are
summarized in the following table.\\
\\
\begin{tabular*}{160mm}{lcc}
 & \hspace*{0.5cm} $\left( e^+ e^- \rightarrow \, \, {\rm hadrons} \right)$ & \hspace*{3cm} Can quarks\\
 Scheme & \hspace*{0.5cm} $\left( e^+ e^- \rightarrow \mu^+ \mu^- \right)$ & \hspace*{3cm} be real? \\ \hline
 
``Quark statistics'' & \hspace*{0.5cm} 2 & \hspace*{3cm} No \\
Para--Fermi statistics & & \\
rank 3 & \hspace*{0.5cm} 2 & \hspace*{3cm} Probably not\\
Nine Fermi--Dirac quarks & \hspace*{0.5cm} 2 & \hspace*{3cm} Yes \\
Han--Nambu, Fermi--Dirac & \hspace*{0.5cm} 4 & \hspace*{3cm} Yes\\
Quark statistics+charm & \hspace*{0.5cm} 10/3 & \hspace*{3cm} No \\
Para--Fermi, rank 3 + charm & \hspace*{0.5cm} 10/3 & \hspace*{3cm} Probably not\\
Twelve Fermi--Dirac + charm & \hspace*{0.5cm} 10/3 & \hspace*{3cm} Yes \\ \hline
\end{tabular*}
\\
\\
\\
\\
In what follows, we shall confine ourselves to the first scheme, as
requiring the last change in the present experimental situation.\\
\newpage
\noindent
{\bf 4. DERIVATION OF THE $\pi^0 \rightarrow 2 \gamma $ AMPLITUDE IN THE
PCAC APPROXIMATION}\\
\\
In the derivation sketched here, we follow the general idea of Wilson's
and Crewther's method. We lean more heavily on the connected light--cone
current algebra, however, and we do not need to assume full conformal
invariance of matrix elements for small values of the coordinate
differences.\\
\\
To discuss the $\pi^0 \rightarrow 2 \gamma $ decay in the PCAC
approximation, we shall need an expression for
\begin{displaymath}
< \, {\rm vac} \mid F_{ea} (x) F_{e \beta}(y)F_{3y} \, ^5 (x) \mid
\, \, {\rm vac} >
\end{displaymath}
\\
when $x \approx y \approx z$. (Here $e$ is the direction in
$SU(3)$ space of the electric charge.) In fact, we shall consider general
products of the form
\begin{displaymath}
< \, {\rm vac} \, \mid F \left( x_1 \right) F \left( x_2 \right)
\cdots F \left( x_n \right) \mid \, {\rm vac} >
\end{displaymath}
where $F's$ stand for components of any of our currents, and we shall
examine the leading singularity when $x_1, x_2, \ldots, x_n$ tend to lie
among a single lightlike line. (The case when they tend to coincide
is then a specialization.)\\
\\
We assume not only the validity of the connected light--cone algebra,
which implies scale invariance for commutators near the light cone, but
also scale invariance for products near the lightcone, with leading
dimension $l = - 3$ for all currents. There may be subtraction terms in
the products, or at least in physical ordered products, for example,
subtractions corresponding to four--dimensional $\delta $ functions in
coordinate space; these are often determined by current conservation.
But apart from the subtraction terms the current products near the
light cone have no choice, because of causality and their consequent
analytic properties in coordinate space, but to obey the same formulas
as the commutators, with $i \pi \varepsilon \left( z_0 \right) \delta
\left( z^2 \right)$ replaced by $\frac{1}{2} \left( z^2 - iz_0 
\varepsilon \right)^{-1}$ for products and $\frac{1}{2} \left( z^2 - i
\varepsilon \right)^{-1}$ for physical ordered products.\\
\\
Our general quantity $< \, {\rm vac} \, \mid F \left( x_1 \right)
F \left( x_2 \right) \cdots F \left( x_n \right) \mid \, {\rm vac} \, >$
may now be reduced, using successive applications of the product formulas
near the light cone and ignoring possible subtraction terms, since all the
intervals $\left( x_i - y_j \right)^2$ tend to zero, as they do when all
the points $x_1$ tend to lie on the same lightlike line.\\
\\
A contraction between two currents
$F \left( x_i \right), F \left( x_j \right)$ gives a singular function
$S \left( x_i - x_j \right)$ times a bilocal $F \left( x_i x_j \right)$.
If we now contract another local current with the bilocal, we obtain\\
$S \left( x_i - x_j \right)S \left( x_k - x_j \right) F \left( x_i, x_k
\right)$ and so on.\\
\\
As long as we do not exhaust the currents, our intermediate states have
particles in them and we are using the connected algebra generalized
to products. Finally, we reach the stage where we have a string of
singular functions multiplied by
$< \, {\rm vac} \, \mid F \left( x_i, x_j \right) F \left( x_k \right)
\mid \, {\rm vac} >$, and the last contraction amounts to knowing the
disconnected matrix element of a current product. However, the leading
singularity structure of this matrix element can also be determined from
the light--cone algebra by requiring consistent reductions of the three
current amplitudes.\\
\\
We can algebraically reduce a three--current amplitude in two possible
ways. For each reduction the algebra implies the existence of a known
light--cone singularity. The reductions may also be carried out for an
amplitude with a different ordering of the currents. One reduction of
this amplitude yields the same two--point function as before, whereas the
other reduction implies the existence of a second singularity in the
two--point function. Hence we may conclude that the leading singularity
of the two--point function when all points tend to a light line is given
by the product of the two singularities identified by these
reductions. Similarly, the leading singularity of the three--current
amplitude is given by the product of the three singularities indicated
by the different reductions. Since the connected light--cone algebra can
be abstracted
from the free quark model, the result of this analysis implies that the
leading singularities of the two-- and three--point functions are also
given by the free quark model (say, with Fermi--Dirac quarks) and the
only undetermined parameter is an overall factor, $N$, which all vacuum
amplitudes must be multiplied.\\
\\
Since the singularity structure of the two--point function is determined,
we can identify at least a part of the leading light line singularity
of the $n$ current amplitudes. Each different reduction of the $n$ current
amplitudes implies free quark singularities associated with this
reduction. For two, three, and four current amplitudes, all of the
singularities can be directly determined from the different reductions.
For the five and higher--point functions not all of the singularities
can be directly determined, but it is plausible that these others also
have the free quark structure.\\
\\
For the asmptotic value of $\sigma \left( e^+ e^- \rightarrow \,
{\rm hadrons} \right) / \sigma
\left( e^+ e^- \rightarrow \mu^+ \mu^- \right)$,
we are interested in the vacuum expected value of the commutator of two
electromagnetic currents, and it comes out equal to $N$ times a known
quantity. Similarly, more complicated experiments testing products of
four currents, for example, $e ^+ e^-$ annihilation into hadrons and a
massive muon pair or $``\gamma '' - ``\gamma ''$ annihilation into
hadrons, might be considered. Also these processes are, in the
corresponding deep inelastic limit, completely determined by the
number $N$.\\
\\
Returning to $\pi^0 \rightarrow 2 \gamma $ in the PCAC approximation, we
have
$< \, {\rm vac} \, \mid F_{e \alpha} (x) F_{e \beta} (y)
F_{3 \gamma} \, ^5 (z) \mid \, {\rm vac} \, >$ as the
three space--time points approach a
lightlike line, apart from subtraction terms, in terms of $N$ times a
known quantity. We now need only appeal to Wilson's argument (as elaborated
by Crewther). The vacuum expected value of the physically ordered product
$T \left( F_{e \alpha} (x), F_{e \beta} (y), \partial_{\gamma}
F_{3 \gamma} \, ^5 (z) \right)$, taken at low frequencies, is what we
need for the
$\pi^0 \rightarrow 2 \gamma $ decay with PCAC, and the Wilson--Crewther
argument shows that it is determined from the small--distance behavior
of $< \, {\rm vac} \, \mid F_{e \alpha} (x) F_{e \beta} (y)
F_{3 \gamma} \, ^5 (z) \mid \, {\rm vac} \, >$,
with the subtraction terms (which are
calculable from current conservation in this case) playing no r\^ole.
This
remarkable superconvergence result, that the low--frequency matrix element
can be calculated from a surface integral around the leading
short--distance singularity (which is the same as the singularity if all
three points tend to a lightlike line), makes possible the derivation of
$\pi ^0 \rightarrow 2 \gamma$ in the PCAC approximation from the
light--cone current algebra. We come out with the Adler result (i. e., the
result for three Fermi--Dirac quarks) multiplied by $N$.\\
\\
Thus the connected light--cone algebra provides a link between the
$\pi^0 \rightarrow 2 \gamma $ decay and the asymptotic ratio
$\sigma \left( e^+ e^- \rightarrow \, {\rm hadrons} \, \right) / \sigma
\left( e^+ e^- \rightarrow \mu ^+ \mu^- \right)$. Of course, one might
doubt the applicability of PCAC to $\pi^0$ decay, or to any process in
which other currents are present in addition to the axial vector current
connected to the pion by PCAC. If the connected algebra is right,
including products, then failure of the asymptotic ratio of the $e^+ e^-$
cross sections to approach the value 2 would be attributed either to such
a failure of PCAC when other currents are present or else to the need for
an alternative model such as we discussed in Section 3.\\
\\
As a final remark, let us mention the ``finite theory approach,'' as
discussed in ref. 4 in connection with the light--cone current algebra.
Here the idea is to abstract results not from the formal ``quark--gluon''
field theory model, but rather from the sum of all orders of perturbation
theory (insofar as that can be studied) under two special assumptions. The
assumptions are that the equation for the renormalized coupling constant
that allows for a finite coupling constant renormalization has a root and
that the value of the renormalized coupling constant is that root. Under
these conditions, the vacuum expected values of at least some current
products are less singular than in the free theory. Since the Adler result
still holds in the ``finite theory case,'' the connected light--cone
algebra would have to break down. In particular, the axial vector current
appearing in the commutator of certain vector currents is multiplied by
an infinite constant. These are at present two alternative possibilities
for such a ``finite theory'':\\
\begin{enumerate}
\item[1.] Only vacuum expected values of products of singlet currents are
less singular than in the free theory;$^{14}$ only the parts of the
algebra that involve singlet currents are wrong (e.g. the bilocal singlet
axial vector current is infinite); the $e^+ e^-$ annihilation cross
section would still behave scale invariantly.
\item[2.] All vacuum expected values of current products are less
singular than in the free theory; the number $N$ is zero; all bilocal
axial vector currents are infinite; the $e^+ e^-$ annihilation cross
section would decrease more sharply at high enegies than in the case of
scale invariance.\\
\end{enumerate}
\bigskip
\bigskip
{\bf 1. ACKNOWLEDGEMENTS}\\
\\
For discussions, we are indebted to D. Maison, B. Zumino, and other members
of the staff of the Theoretical Study Division of CERN. We are pleased to
acknowledge also the hospitality of the Theoretical Study Division.\\
\newpage
\noindent
{\bf REFERENCES}
\begin{enumerate}
\item[1.] H. Fritzsch and M. Gell--Mann, ``Proceedings of the Coral
Gables Conference on Fundamental Interactions at High Energies, January
1971,'' in {\it Scale Invariance and the Light Cone,} Gordon and Breach,
New York, (1971).
\item[2.] H. J. Lipkin and S. Meshkov, {\it Phys. Rev. Letters}, {\bf 14}, 
670 (1965).
\item[3.] R. Dashen and M. Gell-Mann, {\it Phys. Letters}, {\bf 17},
142 (1965). 
\item[4.] H. Fritzsch and M. Gell--Mann, {\it Proceedings of the
International Conference on Duality and Symmetry in Hadron Physics,}
Weizmann Science Press of Israel, Jerusalem, 1971.
\item[5.] G. Zweig, CERN Preprints TH, 401 and 412 (1964).
\item[6.] See, for example, O. W. Greenberg, {\it Phys. Rev. Letters},
{\bf 13}, 598
(1964).
\item[7.] S. L. Adler, {\it Phys. Rev.}, {\bf 177}, 2426 (1969);
J. S. Bell and R. Jackiw, {\it Nuovo Cimento}, {\bf 60A}, 47 (1969).
\item[8.] S. L. Adler, in {\it Lectures on Elementary Particles and
Fields} (1970 Brandeis University Summer Institute), MIT Press, Cambridge,
Mass., 1971, and references quoted therein.
\item[9.] S. L. Adler and W. A. Bardeen, {\it Phys. Rev.}, {\bf 182},
1517 (1969).
\item[10.]  R. J. Crewther, Cornell preprint (1972).
\item[11.] K. G. Wilson, {\it Phys. Rev.}, {\bf 179}, 1499 (1969).
\item[12.] E. J. Schreier, {\it Phys. Rev. D}, {\bf 3}, 982 (1971).
\item[13.] M. Han and Y. Nambu, {\it Phys. Rev.}, {\bf 139}, 1006 (1965).
\item[14.] See also B. Schroer, Chapter 3 in this volume (p. 42).
\end{enumerate}

}
\end{document}